\documentclass[longauth]{aa} 
\usepackage[utf8]{inputenc}
\usepackage[T1]{fontenc}
\usepackage{color}
\usepackage{multirow}
\usepackage{graphicx}
\usepackage{txfonts}
\usepackage[hidelinks]{hyperref}
\usepackage{xcolor}
\hypersetup{
    colorlinks,
    linkcolor={red!50!black},
    citecolor={blue!50!black},
    urlcolor={blue!80!black}
}
\usepackage{gensymb}
\usepackage{wasysym}
\usepackage[flushleft]{threeparttable}

\usepackage{natbib}

\usepackage{footmisc}

\begin{document} 
	
\title{A spectro-polarimetric study of the planet-hosting G dwarf, HD~147513}
	
%	\subtitle{I. Observational Analysis}

\author{
G. A. J. Hussain\inst{1,2}, J. D. Alvarado-G\'omez\inst{1,3}, J. Grunhut\inst{1}, J.-F. Donati\inst{2,4}, E. Alecian\inst{5,6,7}, M. Oksala\inst{7},  
J. Morin\inst{8}, R. Fares\inst{9},  M. Jardine\inst{10}, J.J. Drake\inst{11}, O. Cohen\inst{11}, S. Matt\inst{12},  P. Petit\inst{2,4}, S. Redfield\inst{13}   
%\inst{1} 
	\and	
	F. M. Walter\inst{13}%\fnmsep\thanks{Just to show the usage          %of the elements in the author field}
          }
\institute{\inst{1} European Southern Observatory, Karl-Schwarzschild-Str. 2, 85748 Garching bei M\"unchen, Germany
\email{ghussain@eso.org} \\
\inst{2} Institut de Recherche en Astrophysique et Plan\'etologie, Universit\'e de Toulouse, UPS-OMP, F-31400 Toulouse, France \\
\inst{3} Universit\"ats-Sternwarte M\"unchen, Ludwig-Maximilians-Universit\"at, Scheinerstr. 1, 81679 M\"unchen, Germany \\
 \inst{4} CNRS, Institut de Recherche en Astrophysique et Plan\'etologie, 14 Avenue Edouard Belin, F-31400 Toulouse, France \\
 \inst{5} Univ. Grenoble Alpes, IPAG, F-38000, Grenoble, France \\
 \inst{6} CNRS, IPAG, F-38000, Grenoble, France \\
 \inst{7} 
 LESIA, Observatoire de Paris, CNRS UMR 8109, UPMC, Univ. Paris Diderot,  5 place Jules Janssen, 92190 Meudon, France \\
\inst{8} LUPM-UMR 5299, CNRS \& Universit\'e Montpellier, Place Eug\'ene Bataillon, 34095 Montpellier Cedex 05, France \\
 \inst{9}  INAF-Osservatorio Astrofisico di Catania, Via Santa Sofia 78, 95123 Catania Italy \\
\inst{10} SUPA, School of Physics and Astronomy, University of St Andrews, St Andrews KY16 9SS, UK \\
 \inst{11} Harvard-Smithsonian Center for Astrophysics, 60 Garden Street, Cambridge, MA 02138, USA \\ 
\inst{12} Department of Physics and Astronomy, University of Exeter, Stocker Road, Exeter EX4 4QL, UK \\
\inst{13} Astronomy Department, Van Vleck Observatory, Wesleyan University, 96 Foss Hill Drive, Middletown, CT 06459, USA \\
\inst{14} Department of Physics and Astronomy, Stony Brook University, Stony Brook NY 11794-3800, USA         }

   \date{Received -----; accepted -----}

\abstract{The results from a spectro-polarimetric study of the  planet-hosting Sun-like star, HD 147513 (G5V), are presented here. Robust detections of Zeeman signatures at all observed epochs indicate a surface magnetic field, with longitudinal magnetic field strengths varying between 1.0-3.2 G. 
Radial velocity variations from night to night modulate on a similar timescale to the longitudinal magnetic field  measurements. These variations are therefore likely due to the rotational modulation of stellar active regions rather than the much longer timescale of the planetary orbit ($P_{\rm orb}=528$\,d).
Both the longitudinal magnetic field measurements and radial velocity variations are consistent with a rotation period of 10 $\pm 2$\,days, which are also consistent with the measured chromospheric activity level of the star ($'log R'_{\rm HK} = -4.64$). Together, these quantities indicate a low inclination angle, $i \sim 18^{\circ}$.
We present preliminary magnetic field maps of the star based on the above period and find a simple poloidal large-scale field. Chemical analyses of the star have revealed that it is likely to have undergone a barium-enrichment phase in its evolution because of a higher mass companion. 
Despite this, our study reveals that the star has a fairly typical activity level for its rotation period and spectral type. 
Future studies will enable us to explore the long-term  evolution of the field, as well as to measure the stellar rotation period, with greater accuracy.
}

\titlerunning{Spectro-polarimetric study of HD 147513}
\authorrunning{Hussain et al.}
\maketitle
%
%________________________________________________________________

\section{Introduction}\label{sec_intro}

As efforts to find planets around other stars intensify, there is a need to better characterise stellar magnetic activity on a range of timescales in moderate-activity stars. The HARPS instrument on the ESO 3.6-m telescope at the La Silla Observatory is a high-precision velocimeter, enabling stability to the 1\,m\,s$^{-1}$ level \citep{2005Msngr.120...22Pepe}.
The instrument also offers a polarimetric  mode, which enables the detection of stellar surface magnetic fields \citep{2011Msngr.143....7P}. HARPS is thus ideally suited to both studying stellar magnetic activity and analyse the contribution of magnetic activity to radial velocity ``jitter'' in intensity line profiles (also see \citealt{2011A&A...527A..82Dumusque}). 

{ Spectro-polarimetric observations of cool stars spanning almost three decades have  yielded  not only direct measurements of magnetic fields at the surfaces of these stars \citep{1997MNRAS.291..658DonatiLSD}, but also detailed maps of large scale magnetic field topologies through the technique of Zeeman Doppler imaging (\citealt{1989A&A...225..456Semel},
 \citealt{1991A&A...250..463B},
 \citealt{2009ARA&A..47..333D}). }
Over 40 main sequence G-K stars have been imaged to date, and these studies show general clear trends (see \citealt{2014MNRAS.441.2361Vidotto}). 
The slowest rotators ($P_{\rm rot}>$15\,d) possess the simplest and weakest large scale fields. Fields strengthen and increase in complexity in more rapidly rotating stars. Many of the maps for the most rapidly rotating stars also feature { strong surface  toroidal} fields; a feature that has no clear counterpart on the Sun and very likely indicates a change in the underlying stellar dynamo. 
The BCool collaboration\footnote[2]{Bcool is part of the MagIcS international project - see \href{http://www.ast.obs-mip.fr/users/donati/magics/v1/}{http://www.ast.obs-mip.fr/users/donati/magics/v1/} for more information.}   has collected and analysed circularly polarised spectra of over 170 solar-type stars. { \cite{2014MNRAS.444.3517MarsdenBcool} report magnetic field detections} on 67 of these stars and present these detections in the context of their chromospheric activity, rotation and age. 

We used HARPS in polarimetric mode to obtain high quality S/N spectro-polarimetric time series of two low-moderate activity planet-hosting, solar-type stars. 
These data enable us to  characterise the magnetic activity properties in detail and use these maps to model the coronae and conditions in these young planetary systems. In the first study 
\cite{2015AA...submittedAG} present magnetic field maps of the planet-hosting G dwarf, HD 1237. 
Our time series revealed a rotation period of seven days and the resulting magnetic field maps showed a strong toroidal component. 

We present the results for the second target in the study, HD 147513. 
In Section\,\ref{sec_hd147513} we give a more detailed description of the stellar system. Observations and the analysis of the chromospheric activity are in Sects.\,\ref{sec_data} \&  \ref{sec_rhk}, respectively. 
The photospheric line  profiles are used to measure the longitudinal magnetic field and radial velocity and to study the night-to-night variability in both these quantities. 
Our conclusions are summarised in Sect.\,\ref{sec_summary}, and maps based on our best estimates of the stellar rotation period are presented in the Appendix.

\section{HD 147513}
\label{sec_hd147513}

HD 147513 (GJ 620.1 A, HR 6094) is a bright ($mV=5.4$) G dwarf at a distance of 12.9\,pc \citep{2005ApJS159141ValentiFischer}. 
Its main properties are listed in Table\,\ref{tab_1}; these include both published values and those determined in the analysis presented here. 

HD 147513 is shown to be moderately magnetically active, with an  average X-ray luminosity of about $10^{29}$\,erg\,s$^{-1}$ in the 0.1-2.4\,keV ROSAT PSPC band 
\citep{2004A&A...417..651SchmittLiefke}.
For comparison, the average solar X-ray luminosity is approximately $10^{27.6}$\,erg\,s$^{-1}$, varying by an order of magnitude over the course of the 11-year solar activity cycle \citep{2003ApJ...593..534Judge} 
Chromospheric activity is indicated by significant emission in its Ca II H\&K profiles
with a range of $R'_{\rm HK}$ values reported in the literature ($-4.6$\,$<\log R'_{\rm HK}<$\,$-4.38$) \citep{2005A&A...443..609Saffe}.
The rotation period of the star has been estimated based on the above $\log R'_{\rm HK}$ values, and published estimates range between  4.7 \cite{2004A&A...415..391Mayor} and $8.5$ ($\pm 2.2$\,days; \citealt{2010MNRAS.408.1606Watson}).
Age estimates based on chromospheric activity place the star at 0.45\,Gyr \citep{1998MNRAS.298..332RochaPinto}. The star may also be associated with the  0.5\,Gyr Ursa Major moving group  \citep{2003AJ....125.1980King}.

High precision radial velocity measurements spanning almost five years reveal the presence of a Jupiter-mass planet  
($M\sin i=1.21M_{\rm J}$, $P_{\rm orb}= 528.4$\,d). The semi-amplitude is $K=29.3 \pm 1.8$\,m\,s$^{-1}$ with a dispersion of 5.7\,m\,s$^{-1}$. \cite{2004A&A...415..391Mayor} report that the orbital properties of this planetary system are characteristic of intermediate-to-long orbital period radial velocity planets, with a semi-major axis, $a= 1.32$\,au and eccentricity, $e=0.26$.

As also discussed by \cite{2004A&A...415..391Mayor}, the chemical analysis of HD 147513 points to a  complex evolutionary history. Its high lithium abundance indicates  its relative youth. 
However, the star is also found to be  over-abundant in barium and s-process elements.
\cite{1997ApJ...476L..89Portodemello} suggest that this is due to mass transfer between  HD 147513 and the AGB progenitor of the white dwarf, CD-38$^{\circ}$10980, which is found to have a common proper motion. This well-studied white dwarf has an age of 30\,Myr and is at a distance of 5360 au from HD 147513 with an original mass of 2.6\,$M_{\odot}$. This more massive star may have driven sufficient mass transfer onto  HD 147513 to explain its observed abundance of s-process elements. 
Indeed \cite{1997ApJ...476L..89Portodemello} suggest further that these stars may have been part of a multiple star system, bound with the binary, HR 2047; which is 24\,pc away, a confirmed member of the UMa moving group, and which also shows evidence of barium enrichment (albeit to a lesser extent).

\begin{table}[h]
\caption{HD 147513 basic properties.}             % title of Table
\centering                          % used for centering table
\begin{tabular}{l c c}        % centered columns (4 columns)
\hline\hline                 % inserts double horizontal lines
Parameter & Value & Reference \\
\hline
{ Sp. Type} & G5V & \protect{\cite{1993ApJ...402L...5SoderblomMayor}}\\ 
$B-V$ & 0.62 & \protect{\cite{1993ApJ...402L...5SoderblomMayor}}\\ 
Age [Gyr] & $\sim 0.45$ & \protect{\cite{1998MNRAS.298..332RochaPinto}} \\
$T_{\rm eff}$ [K] & $5930 \pm 44$ & \protect{\cite{2005ApJS159141ValentiFischer}} \\
$\log(g)$ & $4.612 \pm 0.06$ & \protect{\cite{2005ApJS159141ValentiFischer}} \\
$M_{*}$ [M$_{\odot}$] & 1.07$\pm 0.01$ & \protect{\cite{2007ApJS..168..297Takeda}}\\
$R_{*}$ [R$_{\odot}$]& 0.98$^{0.03}_{-0.02}$&\protect{\cite{2007ApJS..168..297Takeda}}\\
$v\sin i$ [km s$^{-1}$] & $1.5 \pm 0.4$  & \protect{\cite{2005ApJS159141ValentiFischer}} \\
$i$ [$^\circ$] & $18^{12}_{-8}$ & This work \\ % 
$v_{\rm R}$ [km s$^{-1}$] & $13.232 \pm 0.09$ & This work \\ 
$P_{\rm rot}$ [days] & $10.0 \pm 2.0$ & This work \\
$\log(R^\prime_{\rm HK})$ & $-4.64 \pm 0.05$ & This work \vspace{2.5pt}\\   
$\log L_X$ & 28.92 & \protect{\cite{2004A&A...417..651SchmittLiefke}} \\
\hline                                   %inserts single line
\end{tabular}
\label{tab_1}      % is used to refer this table in the text
\end{table}

\section{Observations}\label{sec_data}

\noindent In this paper we present high S/N high resolution circularly polarised spectra obtained  using the polarimetric mode of the HARPS echelle spectrograph at the ESO 3.6-m telescope at the La Silla Observatory (\citealt{2011Msngr.143....7P}, \citealt{2003Msngr.114...20M}).

\begin{table}[h]
\caption{Journal of observations. The columns contain the date, the corresponding Barycentric Julian Date (BJD), the start time of the observations in UT, the exposure times, and the Stokes I peak Signal-to-Noise ratio (S/N).}             % title of Table
\label{tab_2}      % is used to refer this table in the text
\begin{threeparttable}
\centering                          % used for centering table
{\small 
\begin{tabular}{l c c  c }        % centered columns (4 columns)
\hline\hline                 % inserts double horizontal lines
Date & BJD (TT) & UT & Stokes I  \\    % table heading 
(2012) & (2400000+) & & Peak S/N  \\    % table heading 
\hline                        % inserts single horizontal line
Jul 15 & 56123.53315 & 00:40:36 &  660 \\     
Jul 15 & 56124.52049 & 24:22:28 &  810 \\
Jul 17 & 56125.54088 & 00:51:56 &  570 \\
Jul 18 & 56126.53188 & 00:39:03 &  720 \\
Jul 18 & 56127.50182 & 23:55:52 &  734 \\
Jul 20 & 56128.53529 & 00:44:09 &  923 \\
Jul 22 & 56130.59443 & 02:09:32 &  550 \\
Jul 23 & 56131.71964 & 05:09:56 &  340 \\
\hline  %inserts single line
\end{tabular}}
\end{threeparttable}
\end{table}

The spectra encompass a wavelength range  from 378 nm to 691 nm, with a 8\,nm gap centred at 530\,nm. The data were acquired in 2012 July under changeable weather conditions and a full observation log is shown in Table\,\ref{tab_2}. All the exposures had the same exposure time of 3600\,s for the full circularly polarised spectrum (Stokes V) sequence. This is obtained by combining four individual sub-exposures using the ratio method (see \citealt{1997MNRAS.291..658DonatiLSD, 2009PASP..121..993B}) and enables a null-polarisation spectrum to be constructed in order to check for possible spurious polarisation contributions to the Stokes V profiles \citep{1997MNRAS.291..658DonatiLSD}. 

Data were reduced using the ESPRIT package which has been adapted for the HARPS instrument (\citealt{1997MNRAS.291..658DonatiLSD}, H\'ebrard et al. {\em {in prep.}}). 
This package produces an optimal extraction of the bias-subtracted spectra after flat-fielding corrections. 
{ The slit shape is averaged over each order and used to compute the curvilinear coordinate system along which the spectra are extracted. The calibration frames required by the package are the bias frames, flat field frames and a good quality ThAr arc spectrum that were acquired each night. Spectra extracted using the REDUCE package \citep{2002A&A...385.1095P,2011A&A...525A..97M} were almost identical compared to those reduced with the ESPRIT, with the latter  showing slightly higher S/N levels.
As barycentric corrections are also applied to the spectra reduced by ESPRIT, this is the dataset used in the analysis presented here.} 
The extracted data have spectroscopic resolutions varying from 95\,000 to 113\,000, depending on the wavelength, with a median value of 106\,000. Given the noise level of these data 1\,m\,s$^{-1}$ accuracy should be achievable in the radial velocity measurements. 

\begin{figure}[ht]
\centering %  left, bottom, right and top
\includegraphics[trim=0.0cm 0.8cm 25cm 1.4cm, clip=true, width=\hsize]{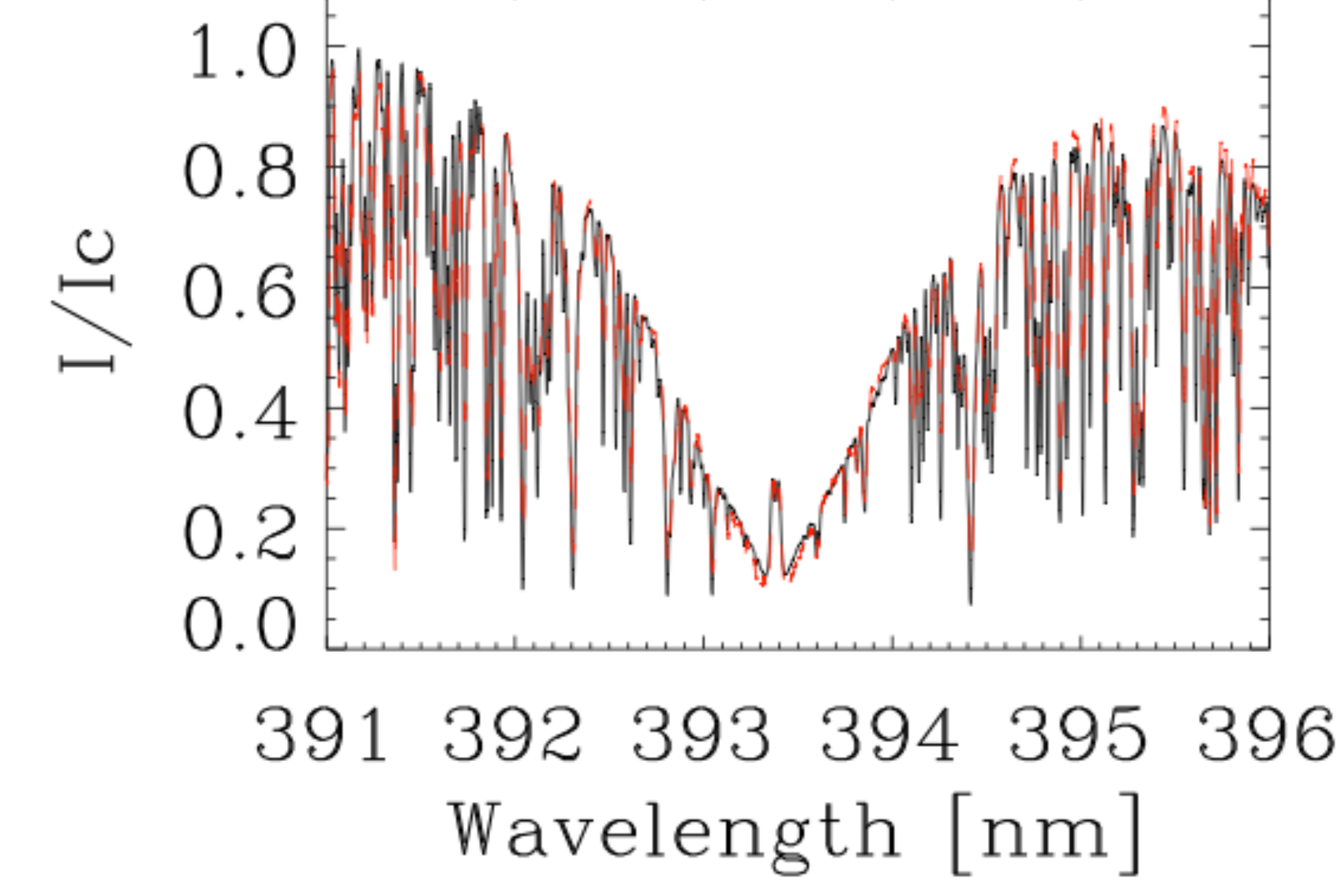}
\caption{Chromospheric activity in HD 147513: Comparison between Ca II K observed with HARPS in 2012 July 18 (red dotted line)  and archive FEROS spectra from 2006 July 15 (black solid line). 
The HARPS spectra have been rebinned to enable a better comparison with the FEROS spectra, which have a lower spectroscopic resolution.}
\label{fig_1}
\end{figure}

\section{Chromospheric activity}\label{sec_rhk}

In order to characterise the chromospheric activity level of the star at the observed epoch, we present an analysis of the Ca\,II H (396.8492 nm) \& K (393.3682 nm) lines, converting the fluxes to the classic Mount Wilson S-index, $S_{\rm MW}$. This index is defined as follows:

\begin{equation}\label{eq_1}
S_{\rm MW} = \dfrac{H + K}{R + V}\mbox{ .}
\end{equation}

\noindent Here $H$ and $K$ represent the fluxes measured in each of the Ca II line cores using $0.105$ nm wide spectral windows. $R$ and $V$ are the fluxes measured in the continuum over $2$ nm windows centred at $390.1$ nm and $400.1$ nm respectively, on both sides of the Ca II region. 
The HARPS S-index was converted to the  Mount Wilson scale by first measuring a calibration factor, $\alpha$, for a series of standard stars observed with HARPS that also have published $S_{\rm MW}$ values. 
{ The spectra from a number of G and K-type stars were renormalised to the continuum level in the same way and aligned using a high S/N HARPS solar spectrum as a template. The largest contribution to the measurement errors is likely due to slight differences in the continuum normalisation for each individual exposure. We find that the renormalisation introduces a typical error of 5\%.  The errors on the values of the S-index are however dominated by the conversion factor, $\alpha$, and we  refer to  \cite{2015AA...submittedAG} for further details.} Their  conversion factor,  $\alpha = 15.39 \pm 0.65$, is used to compute the HARPS S-index using the measured HARPS fluxes, $H$, $K$, $R$, and $V$:

\begin{equation}\label{eq_2}
S_{\rm MW} = 
\alpha\left(\dfrac{H + K}{R + V}\right)_{\rm H}%
\end{equation}

Sample spectra from our dataset are shown in Fig.\,\ref{fig_1} (red line), clearly illustrating emission in the cores of both Ca\,II H\&K profiles due to significant chromospheric heating and indicating a moderate magnetic activity level. An average S-index of 0.23 $\pm 0.01$ 
is computed for our dataset. No significant variability is found in the Ca II H\&K fluxes over eight days, indicating a constant contribution from the chromospheric active regions even as the star rotates
(with estimated rotation periods in the literature ranging from 4.7--8.5\,d). 

The S-index is converted to the chromospheric activity indices, $R_{\rm HK}$ and $R^\prime_{\rm HK}$  applying the colour and photospheric corrections for main sequence cool stars and a $B-V$ of 0.62 \citep{1982A&A...107...31M,1984ApJ...279..763Noyes}. 
This conversion results in an average $\log R'_{\rm HK}$ of -4.64 $\pm 0.06$ (from $\log R_{\rm HK}= -4.44 \pm 0.04$). 

As noted in Sect.\,\ref{sec_hd147513} a wide range of $\log R'_{\rm HK}$ values in the literature, ranging from $-4.6$ to $-4.38$ over a period spanning almost 20 years (from 1983 to 2004;  \citealt{1987AJ.....93..920Soderblom}, \citealt{2005A&A...443..609Saffe}).
In order to investigate whether long-term changes (e.g., due to magnetic activity cycles) might be at the root of these different measurements we searched public archives for spectra of HD 147513 that cover the relevant wavelength range. Fig.\,\ref{fig_1} shows a comparison between spectra acquired in 2006 (archive FEROS spectra) with our 2012 HARPS spectra. It is clear that there is little variability over these two epochs and the corresponding $R'_{\rm HK}$ indices are therefore identical within the measurement errors. 
While this cannot exclude intrinsic variability over a wider range of timescales it is likely that the chromospheric activity level is more stable than suggested from the range of published measurements. We conclude that these variations are likely dominated by differences in conversions to the Mt Wilson index from spectra acquired from a range of instruments.

It is possible to estimate the rotation period of the star within about 20\% accuracy using its $\log R'_{\rm HK}$ index and the conversion factors presented by \cite{1984ApJ...279..763Noyes}. The Rossby number of the star is computed using its $R'_{\rm HK}$ index, while the convective turnover timescale, $\tau_c$ can be estimated from the star's $B-V$. 
We find a period of 12.4\,d, for our value ($\log R'_{\rm HK}=-4.64$); this is somewhat larger but still compatible with the 8.5-d estimate (\citealt{2010MNRAS.408.1606Watson}; based on  a $\log R'_{\rm HK}$ of $-4.52$). However, it is completely incompatible with the 4.7-d value based on the highest $\log R'_{\rm HK}=-4.38$ \citep{2004A&A...415..391Mayor}. Combining these estimates for the stellar rotation period with its projected rotational velocity, $v_e \sin i$, and radius (Table\,\ref{tab_1}), it is possible to compute the inclination angle of the star. For HD 147513 a relatively low inclination angle is expected; using the range of 4.7--12.4\,d periods and radius (Table\,\ref{tab_1}) the star's inclination angle must be between  10--25$^{\circ}$.

\section{Photospheric line profiles  \& stellar magnetic field}\label{sec_mag_signatures}

\noindent As the large scale magnetic field in cool stars such as HD 147513 is expected to be relatively weak ($\ll$1\,kG), it is not possible to detect significant polarisation in individual photospheric line profiles. It is therefore necessary to employ a  multi-line technique, e.g. Least Squares Deconvolution (LSD, \citealt{1997MNRAS.291..658DonatiLSD}), to exploit the full wavelength coverage of the dataset (378--691\,nm) and use the signal from thousands of photospheric spectral lines. 
It is typically possible to enhance the  { S/N by a factor of $\sim 30$}, compared to the original spectrum in this way. 

The mask used in the LSD analysis is constructed from an atomic line list extracted for a star with the same basic parameters ($T_{\rm eff}$, $\log g$) as HD 147513 from the VALD database\footnote[2]{\url{http://vald.astro.uu.se/} -- Vienna Atomic Line Database (VALD3)} \citep{2000BaltA...9..590K}. 
This downloaded line list is first ``cleaned'' of all strong lines, including any diagnostics that are likely to have significant contributions from the chromosphere (e.g., Ca II H\&K, H$\alpha$). This list is then further tailored to the star by  adjusting the individual depths to fit those of the spectral lines of HD 147513. As discussed by \cite{2015AA...submittedAG} this ``clean-tweaking'' method (see \citealt{2012MNRAS.426.2738Neiner}) is  most commonly employed when { applying LSD to hot (OB) stars. }
As cool stars have thousands of photospheric line profiles, this technique does not have as significant an impact on the LSD profiles but does  increase them in S/N by between 5-10\% compared to the original ``clean'' line mask
\citep{2015AA...submittedAG}. LSD is then applied to our spectro-polarimetric dataset using these tailored clean-tweaked masks, cutting off at a depth of 0.1. This results in almost 4500 lines being used in the deconvolution. {The velocity step used is 0.8 km/s, which corresponds to the average pixel size of the CCD.}

\subsection{LSD profiles}\label{LSD_profiles}

\begin{figure}[h]
\centering %  left, bottom, right and top
\includegraphics[width=\hsize]{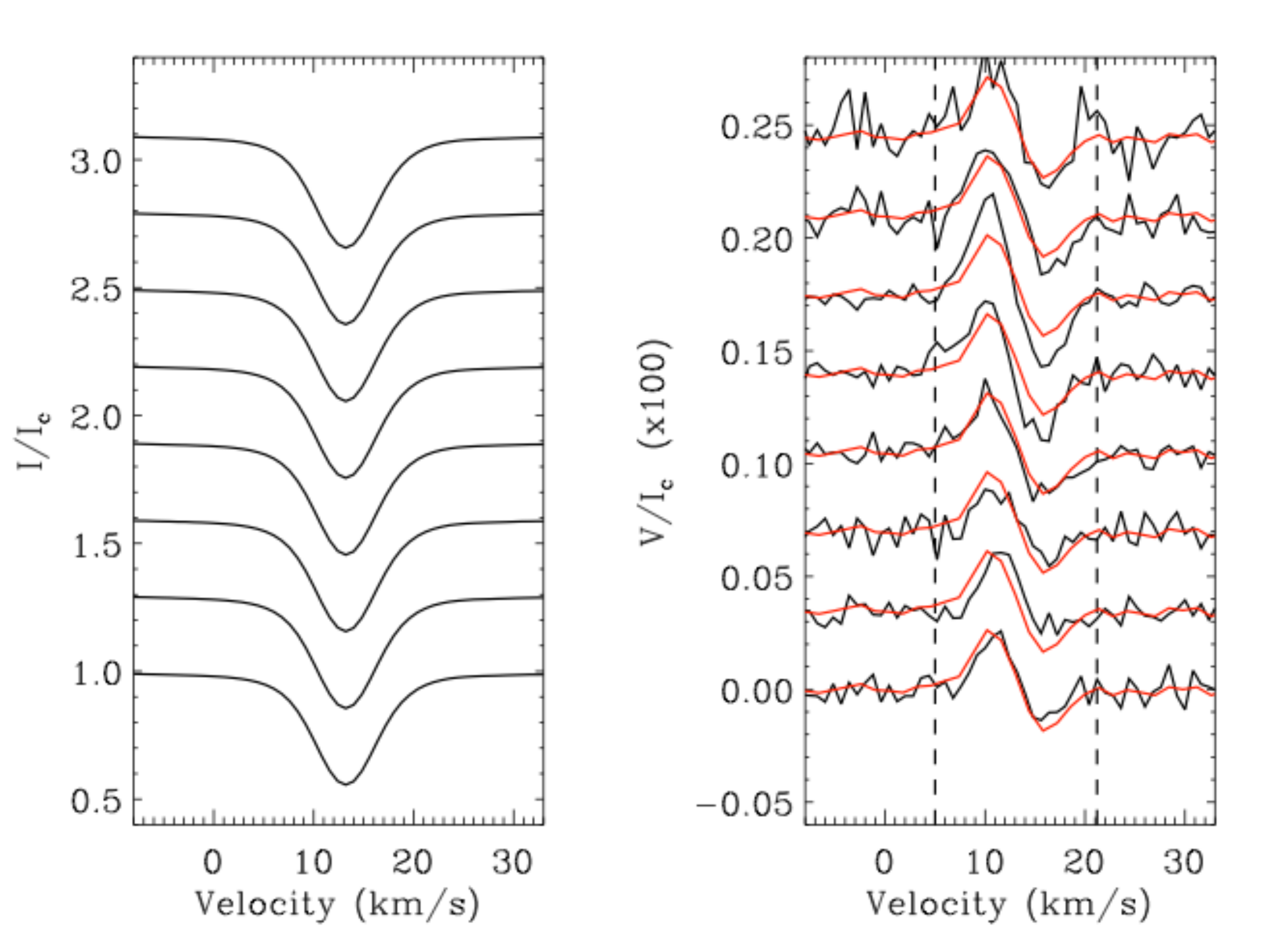}
\caption{LSD Profiles of HD 147513. {\em Left:} Stokes I (intensity) and {\em right:} Stokes V (circularly polarised)  profiles. The mean Stokes V profile computed over this dataset has been overplotted (red line) to investigate variability from night to night. The dashed vertical lines denote the velocity limits over which the 
 $B_\ell$ measurements were calculated.}
\label{fig_5}
\end{figure}

Fig.\,\ref{fig_5} shows the time series of the derived LSD profiles of HD 147513 over 8 days.The Stokes I (unpolarised) profile is shown in the left column, while the Stokes V (circularly polarised) profiles for each epoch are compared to the mean profile on the right. %
The noise level in the Stokes I LSD profiles remains fairly constant ($\sim$9.8\,10$^{-4}$) over the whole dataset. Definite positive magnetic field detections are found in each of the circularly polarised (Stokes V) profiles. From Fig.\,\ref{fig_5} it is clear that the shape of the Stokes V profiles is largely unchanged over the course of the observations, showing a classic antisymmetric shape with respect to the centre. There does however, appear to be a modulation in the amplitude of the Stokes V profiles which is indicative of a small level of inhomogeneity and non-axisymmetry in the large scale field of the star.

\subsection{Longitudinal magnetic field}\label{B_los}

 The derived LSD profiles can be used to compute the  surface averaged longitudinal magnetic field ($B_{\ell}$). This quantity is measured with respect to the intensity line profile, using the central wavelength $\lambda$ (0.519 $\mu$m) and the mean Land\'e factor, $\bar{g}$ (1.197) of the LSD profiles according to the following formula \citep{1997MNRAS.291..658DonatiLSD,2000MNRAS.313..851W}. 

\begin{equation}\label{eq_9}
B_{\ell} = -714 \dfrac{\int v \mbox{V} (v) dv}{\lambda\bar{g}\int \left[1 - \mbox{I} (v)\right] dv} \mbox{ ,}
\end{equation}

\begin{figure}[!ht]
\centering %  left, bottom, right and top
\includegraphics[width=\hsize]{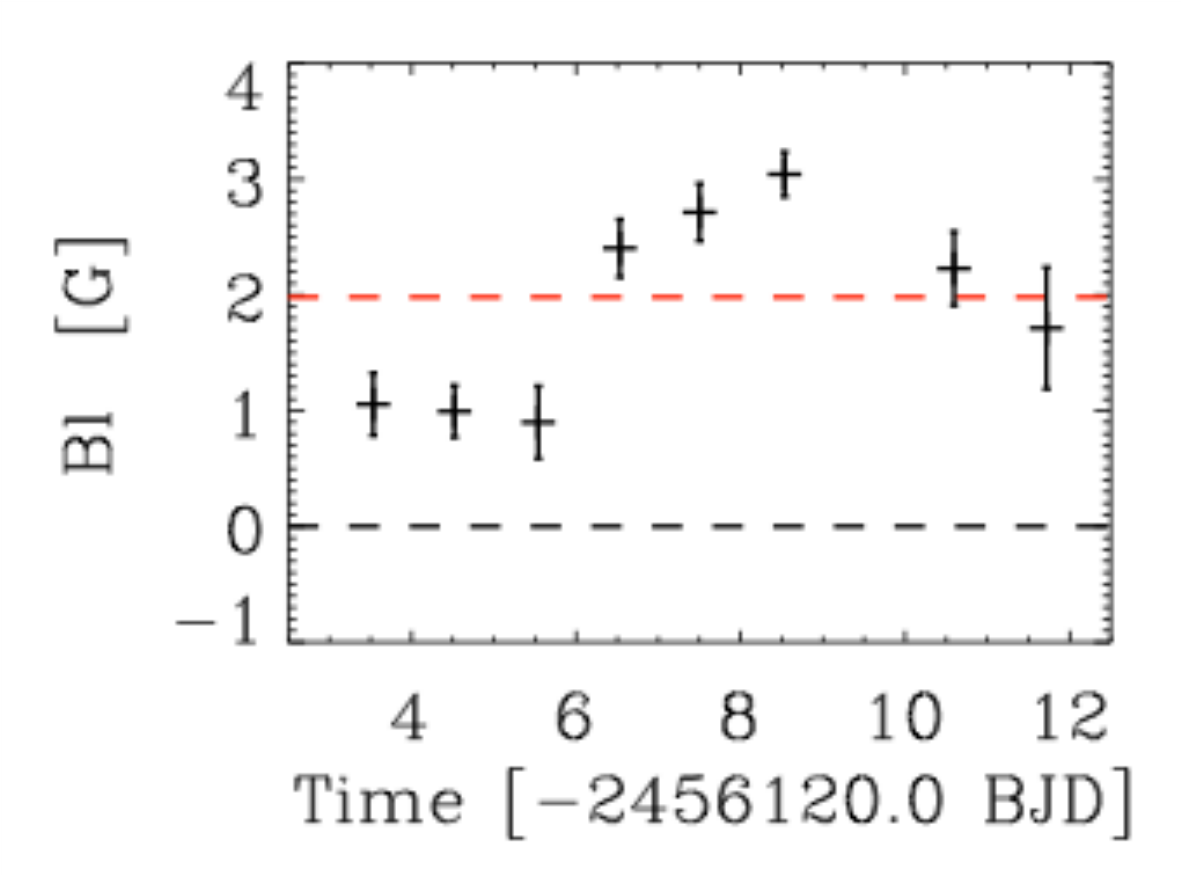}
\caption{$B_{\ell}$ measurements of HD 147513 show significant variability over the 8-day span of the dataset. The x-axis shows the time in Barycentric Julian Date (see Table\,\ref{tab_2}) and the red and black dashed lines denote the mean $B_{\ell}$ (2.0\,G) and  0\,G levels respectively. }
\label{fig_6}
\end{figure}

The measurements of $B_\ell$ for HD 147513 are calculated between 5.0 and 21.2\,km\,s$^{-1}$ from the line centre and show  variability from night to night. {The uncertainties on these values are determined via standard error propagation from the spectra.}
{ The range of values shown in Fig.\,\ref{fig_6} is higher than the range typically seen on the Sun, where $|B_{\ell}|$ is predominantly under 1 G;  solar $|B_{\ell}|$ values can get as high as 3-4G but only very rarely \citep{1998ApJS..116..103K}. }
As noted in Sect.\,\ref{sec_intro}, the chromospheric and coronal magnetic activity levels of the Sun and HD 147513 are very different. 
It is therefore highly likely that these surface magnetic field measurements have a different origin; HD 147513 should have much larger, stronger active regions at the stellar surface compared to the Sun. The $B_{\ell}$ modulations observed in Fig.\,\ref{fig_6} are significant (over 3\,$\sigma$) and their timescale is consistent with that expected by rotational modulation of active regions.
Fig.\,\ref{fig_6} strongly discounts the possibility of a period shorter than 8\,d and hence excludes the previously published value of 4.7\,d.  
Naturally this argument assumes that the variability is not driven by the emergence of new flux. { Studies of active cool stars tracing starspot lifetimes typically show that the large scale field should remain stable over a period  of several weeks and so this appears to be a reasonable assertion (\citealt{1998MNRAS.299..904B}, \citealt{2002AN....323..349H}, \citealt{2009A&ARv..17..251S}). } %

\begin{figure}[!ht]
\centering %  left, bottom, right and top
\includegraphics[width=\hsize]{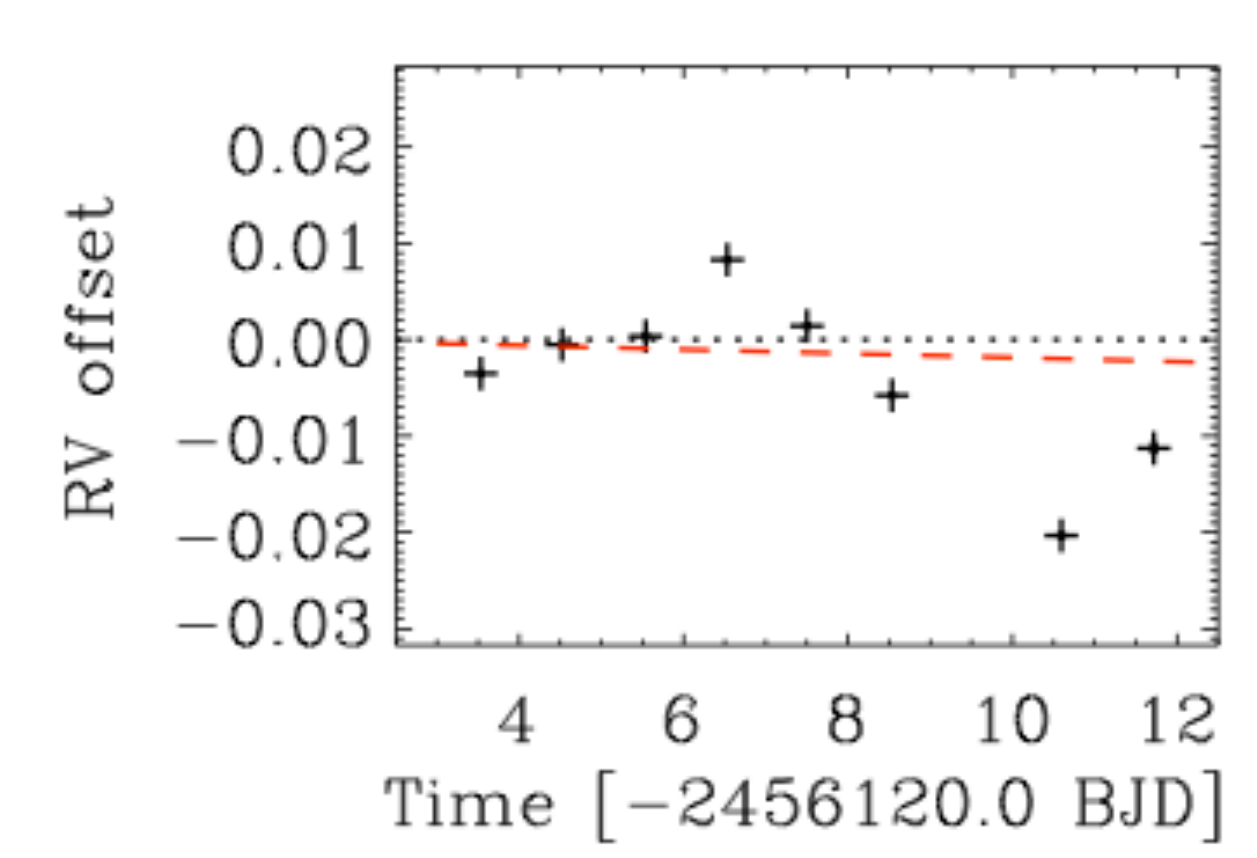}
\caption{RV variability in m\,s$^{-1}$ over the 8-day span of the dataset about the mean (dotted line -- 13.232\,km\,s$^{-1}$). The red line shows the predicted RV contribution of the planet over the same timeframe using the reported ephemeris \citep{2004A&A...415..391Mayor}.
As in Fig.\,\ref{fig_6},  the x-axis shows the time in Barycentric Julian Date. }
\label{fig_7}
\end{figure}

\subsection{Radial velocity and activity}\label{rv_signatures}

We measured the radial velocity (RV) at each observation epoch by least-squares fitting Gaussians to the Stokes I LSD profiles. 
The resulting measurements are shown in Fig.\,\ref{fig_7}, the errors are smaller than the symbol sizes ($\pm 1$\,m\,s$^{-1}$). 
We find significant variations about a mean RV, $\bar{v}=12.232 \pm 0.009$\,km\,s$^{-1}$ over 8-days.  
As the span of our observations is so much smaller than the orbital period of the planet it is clear that these variations are stellar in origin.This is  clearly demonstrated in the comparison of the measured RV variability with the expected RV contribution caused by HD147513b in Fig.\,\ref{fig_7}. This was computed using the ephemeris reported by  \cite{2004A&A...415..391Mayor}.
We note that the timescale of these variations is consistent with that shown by the $B_{\ell}$ measurements (Fig.\,\ref{fig_6}). 
Unfortunately, as less than one full rotation period is sampled in these observations, it is not possible to establish whether this is definitively due to rotational modulation of active regions; although this  appears to be the most likely explanation. 
In particular, as these spectra were obtained by integrating over 1\,hour, shorter timescale phenomena (e.g., pulsations, granulation) are unlikely to contribute significantly to these RV measurements
\citep{2011A&A...527A..82Dumusque}.

\section{Discussion and conclusions}\label{sec_summary}

We have presented an analysis of spectro-polarimetric data of the planet-hosting barium-rich G dwarf, HD 147513. We measured an S-index of 0.23 using the Ca II H\&K lines and use this to compute a mean chromospheric activity index, $\log R'_{\rm HK}$, of $-4.64$.
We compared our HARPS spectra with archive FEROS Ca II H\&K spectra acquired six years previously and find an identical level of chromospheric emission in the cores of the profiles, which indicates that the level of the activity remains more constant than the range of published $\log R'_{\rm HK}$ indices would suggest \citep{2005A&A...443..609Saffe}.

We obtain robust magnetic field detections at all observed epochs and note that this star belongs to the group of low-moderate activity  stars as classified in the ``Bcool'' project.  \cite{2014MNRAS.444.3517MarsdenBcool} find that stars with a similar S-index typically have a 60\% chance of a { definite magnetic field detection}. In the same paper they conclude there is a smaller, 40\% chance, of detecting a magnetic field in stars with similar $v_e \sin i$ values ($<2$\,km\,s$^{-1}$). The activity measurements for HD 147513 ($\log R'_{\rm HK}$ of $-4.64$ and  longitudinal magnetic field values, $0.96<B_{\ell}<3.2$\,G) reported here indicate that  it is fairly typical compared to the BCool star sample  (Fig.\,15 in \citealt{2014MNRAS.444.3517MarsdenBcool}).
HD 147513 falls in the middle of the corresponding activity bin of the BCool sample, indicating that as with other cool stars its activity is determined predominantly by its age and rotation despite its unusual evolution and barium enrichment (see Sect.\,\ref{sec_hd147513}).

Stellar rotation periods can be estimated to about 20\% accuracy based on their chromospheric  $\log R'_{\rm HK}$ index. Our measurements indicate a period of 12.4\,days. Least-squares fitting of sine-curves to the $B_l$ and RV measurements  (Figs.\,\ref{fig_6} \& \ref{fig_7}) reveal rotation periods of 10.4\,d and 9.3\,d respectively, though longer periods cannot be excluded. Fitting both together we find a rotation period of 10\,d, which is the period adopted for the reconstruction in the Appendix. We can therefore definitively exclude the 4.7\,d reported by \citep{2004A&A...415..391Mayor}.  A 10-d period is consistent with  an age of $\sim$0.5\,Gyr using different activity saturation-threshold braking laws (\citealt{1997ApJ...480..303Krishnamurthi}, \citealt{2012ApJ...746...43Reiners}). 
However, as large spreads are found in rotation periods of stars with similar masses at these ages, it is simply noted that the gyrochronological age derived for HD 147513 is consistent with the age of the UMa group.

Significantly longer periods ($>20$\,d) are ruled out due to the relatively high values of the chromospheric and coronal activity indices. The low $v_e \sin i$ of 1.5\,km\,s$^{-1}$ is therefore likely due to a low inclination angle. With a period of 10\,d we compute an inclination angle of 18$^{\circ}$, which is in good agreement with  $15^{+8{\circ}}_{-6}$, as determined by \cite{2010MNRAS.408.1606Watson}. { Based on the maximum and minimum likely values of the radius, rotation period estimates and vsini measurements we find that the inclination can vary between 10 to 30$^{\circ}$. }

Even though a good quality time series of HD 147513 was acquired, as the period covered appears to be less than the rotation period of the star, we cannot definitively measure the rotation period of the star using ZDI as done in previous studies (recent examples; \citealt{2014A&A...569A..79Jeffers}, \citealt{2015AA...submittedAG}). 
A map of the large-scale surface magnetic field is produced using our best estimate of the period and the technique of ZDI in the Appendix (Fig.\,\ref{fig_maps}). We investigate how the large scale structure is affected by the rotation period, varying the period between 8--12\,days, and find that some aspects of the large scale field change (e.g., the magnetic field strength and energy). Regardless of the rotation period used no significant toroidal component is required and the Stokes V signatures can be adequately fit assuming a purely poloidal field. This is different to the ZDI analysis of the G8V star, HD 1237 ($P_{\rm rot}=7$\,d); which shows a dominant toroidal field (\citealt{2015AA...submittedAG}).
While a strong toroidal field of the type detected on HD 1237 can be definitively excluded in HD 147513, the reconstruction is not very sensitive to weaker toroidal field components. Furthermore, as shown in other stars with similar or slightly higher activity levels, e.g., $\epsilon$ Eri and $\xi$ Boo, the toroidal field component may have a stronger contribution at other epochs (\citealt{2014A&A...569A..79Jeffers}, \citealt{2012A&A...540A.138M}).

We computed the expected RV signature of the planet and found this to be almost constant over the 8\,d timescale probed by our observations. As noted, the measured RV variations show a similar modulation  to that traced by $B_{\ell}$ and are consistent with the same period.
It is therefore likely that the RV variability is stellar in origin and due to magnetic activity, e.g., due to a dark spot aligned with the dipolar field. Both surface spots and plage and broadening effects due to the { small scale local field } are found to affect the shape of the line profiles albeit in different ways (e.g., \citealt{2014ApJ...796..132Dumusque}, \citealt{2014MNRAS...443...2599Hebrard}). 

We calculate a RMS of 9\,m\,s$^{-1}$ in these RV measurements. This is 50\% larger than the reported $\sigma(O-C)$ of $\pm 5.7$\,m\,s$^{-1}$  \citep{2004A&A...415..391Mayor}. Those  measurements were based on 30 observations acquired over 1690\,d whereas ours have been collected on a timescale closer to the star's rotation period. The $K$-velocity amplitude due to the planetary orbit is $29.3 \pm 1.8$\,m\,s$^{-1}$. This is a  higher level of activity jitter than previously reported and further observations would be necessary to confirm whether this level of jitter is typical for the star but would likely not significantly affect the planet detection.

We note that the RV RMS we measure is of the same order as that reported in the moderately active M2.5 dwarf, GJ 674,  by \cite{2007A&A...474..293B}.
Whereas the spot causing the RV variability in GJ 674 shows a clear correlation with chromospheric and photospheric spectral indices, no such correlation is found for HD 147513. In HD 147513, the chromospheric activity index remains constant over the eight-day span of the observations. This different relationship between the RV signature of the active region and the chromospheric activity index may be due to the different spectral types of the stars, the lower chromospheric activity index of HD 147513 or differences in the geometric properties of the spot signatures in these two stars. Further studies combining spectro-polarimetry with velocimetry are necessary to better understand the dependence of RV jitter on these parameters.

\begin{acknowledgements}
\noindent Based on observations made with ESO Telescopes at the La Silla Paranal Observatory under the programme ID 089.D-0138 and using spectra downloaded from the ESO Science Archive Facility under the request number GHUSSAIN-162114. We also thank the IDEX initiative at Universit\'e F\'ed\'erale Toulouse Midi-Pyr\'en\'ees (UFT-MiP) for funding the ``STEPS'' collaboration through the Chaire d'Attractivit\'e programme, which enables GAJH to carry out regular research visits to Toulouse.
\end{acknowledgements}

\begin{appendix}
\section{Magnetic field maps of HD 147513}
We present here the maps of the large scale surface magnetic field of HD 147513 assuming the  10-d rotation period that provided the best fit to the longitudinal field and RV variability reported earlier in the paper. 
These have been reconstructed using the ZDI code presented in \cite{2002ApJ...575.1078Hussain}; this describes the field in terms of spherical harmonics and allows for both poloidal and toroidal field components.
{
The local line profile has been modelled using a Milne-Eddington profile whose width and amplitude were adjusted to fit that of HD 147513 and the equivalent width of 67\,m\AA, following the approach of  \cite{2008MNRAS.390..545D}. 
{ More specifically a Voigt profile was used in order to better fit the wings of the Stokes I LSD profiles in this low $v_e \sin i$ star. The width of the local profile was adjusted to find the best fit to the integrated Stokes I profile in agreement with the published $v_e \sin i$ value. }
The model fits shown here all assume a linear limb darkening law, with a limb darkening coefficient of 0.65 \citep{2010A&A...510A..21S}.}

\begin{figure}[!ht]
\centering %  left, bottom, right and top
\includegraphics[width=\hsize]{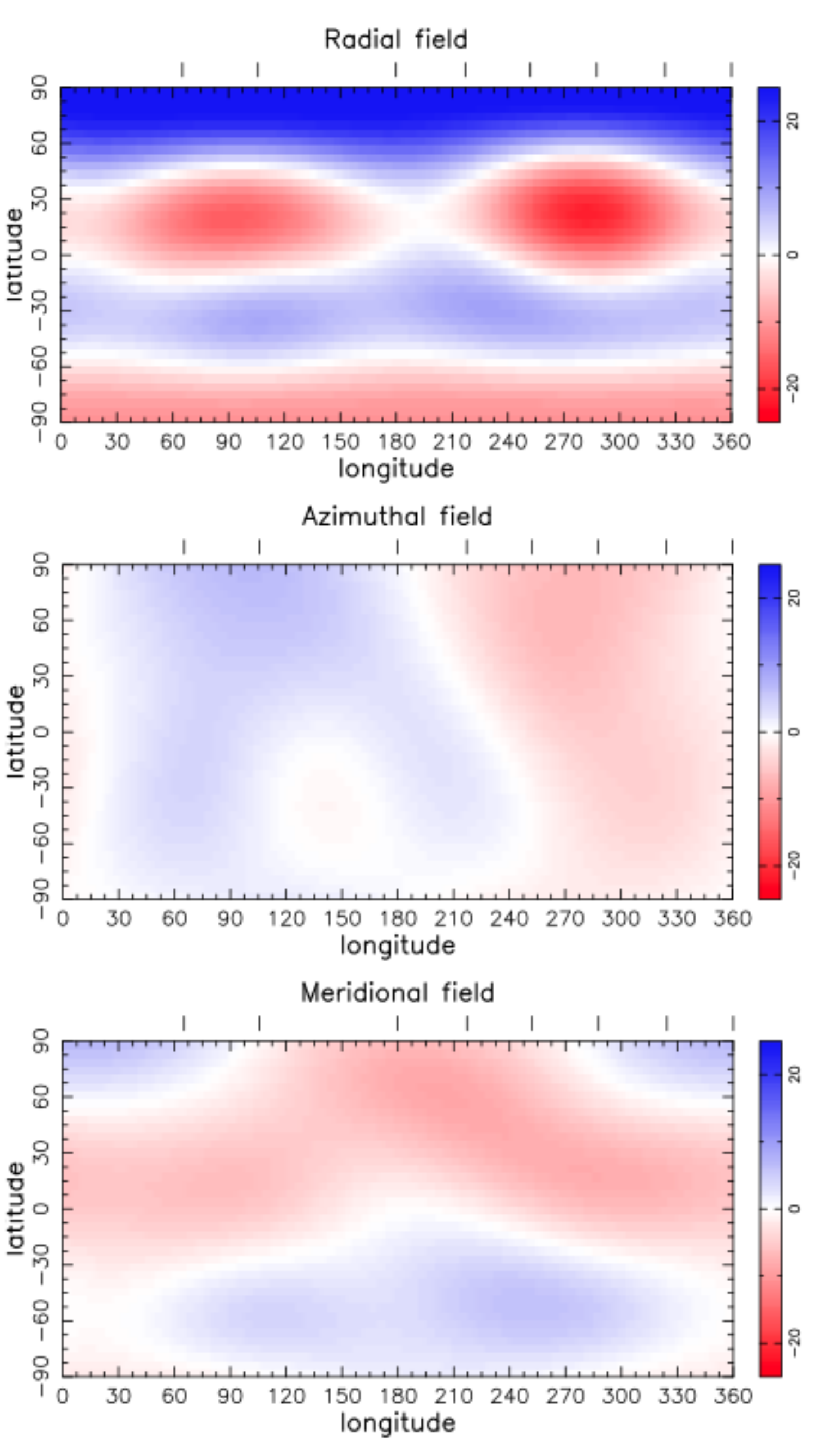}
\includegraphics[width=\hsize]{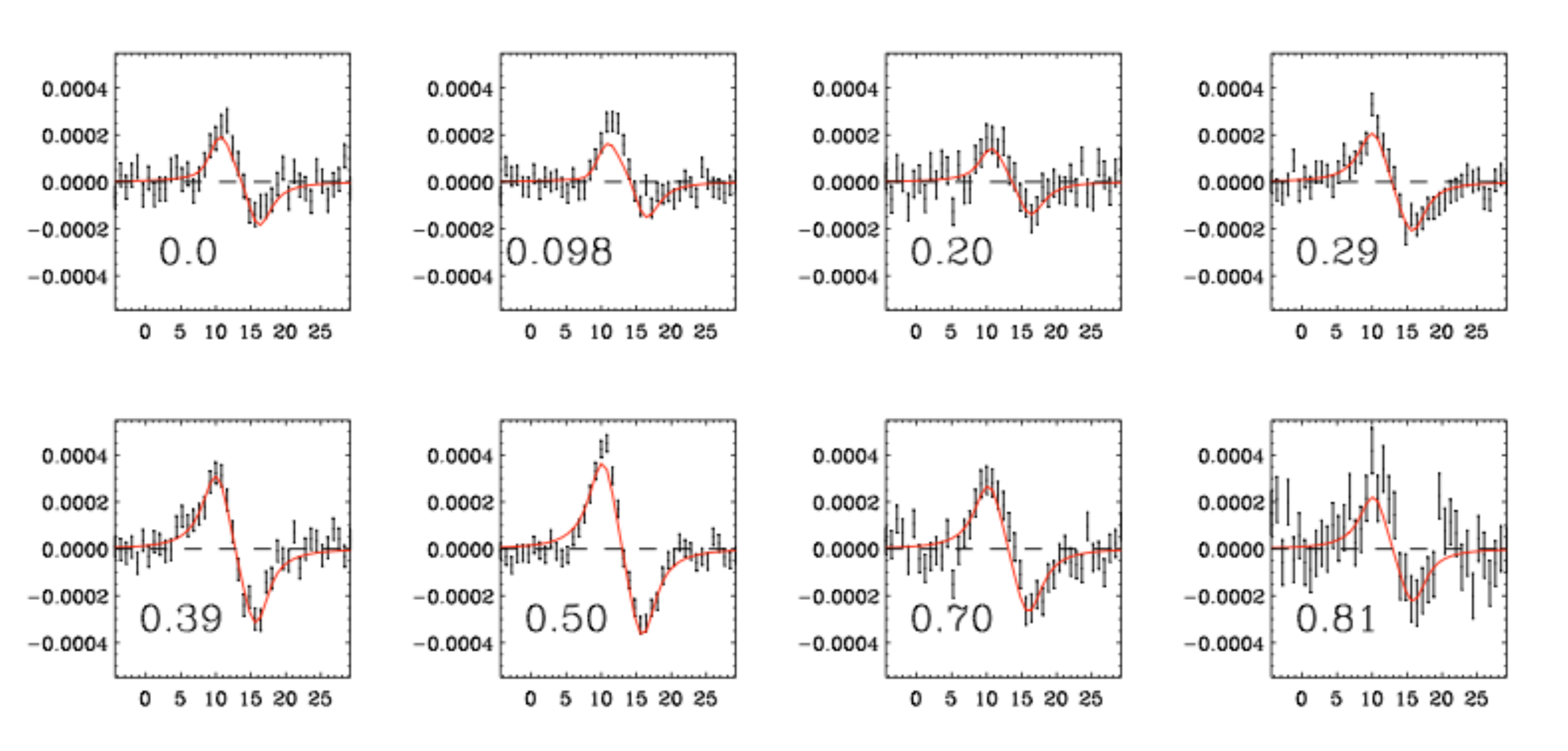}
\caption{Magnetic field maps and fits to the Stokes V data. The first three panels show the radial, azimuthal and meridional field components respectively. Red and blue represents $\pm 25$\,Gauss. The bottom panel shows the fits to the Stokes V profiles ($\chi^2_r=1$). The corresponding rotation phases are listed in the bottom left corner assuming a 10-d rotation period.}
\label{fig_maps}
\end{figure}

The resulting maps fit the observed data to a reduced $\chi^2$ of 1 and are shown in Fig.\,\ref{fig_maps}. { By restricting the solution to a dipolar solution ($l_{\rm max}= 1$) convergence cannot be found beyond $\chi^2_r$ of 1.4. We adopt a maximum spherical harmonic degree of $l_{\rm max}= 3$ as  we find no significant improvement in the fit by allowing higher-order modes. }
These maps have  been derived assuming an inclination angle of 18$^{\circ}$.
Magnetic field regions down to $-18^{\circ}$ latitude should contribute to the observed line profiles in stars with inclination angles of 18$^{\circ}$.The region below the equator is mostly not visible due to the low inclination angle. Nevertheless field is reconstructed in the unobserved hemisphere due to the low order spherical harmonics used. { Most of the energy (70\%) is concentrated in the aligned dipolar, quadrupolar and octupolar components (25\%, 24\% and 21\% respectively), with the rest predominantly divided between the $l=1,m=1$ and $l=2,m=2$ modes.}

We note that no extra toroidal field component was required to fit the data (also see Sect.\,6). Toroidal fields have weaker contributions to circularly polarised profiles in low inclination angle stars as Stokes V profiles are sensitive to the  line-of-sight component of the magnetic field. Hence toroidal fields, particularly at high latitudes will have a weak contribution compared to radial field regions with similar strengths. Despite this it is still possible to exclude the presence of the type of dominant unidirectional azimuthal feature recovered in the maps of the more rapidly rotating G-type stars, HD 1237 and $\xi$ Boo  \citep{2015AA...submittedAG} and  \citep{2012A&A...540A.138M} as the signature would still be unambiguously detectable.  $\xi$ Boo is itself a relatively low inclination star ($i=28^{\circ}$), and shows a dominant toroidal field component at its highest activity states. Future observations of HD 147513 are necessary to reveal whether or not there is a similar change in the relative strength of the toroidal field component with its activity level.

We analysed how the uncertainty in the rotation period may affect the large scale field by 
reconstructing maps assuming periods of 8 and 12\,days. 
The  general structure remains very similar to that shown in Fig.\,\ref{fig_maps} with the main features being either concentrated or smeared out with the shorter and longer period.
The main difference is in the strength of the magnetic flux, which is 20-25\% stronger 
and weaker in the maps derived for the 8-d and 12-d periods respectively. This is expected as the phase coverage is more sparse in the 12-d map. 
Further observations spanning 16--20\,days would be necessary to  ascertain the period with greater accuracy.

\end{appendix}
%-------------------------------------------------------------------

\bibliographystyle{aa}
\bibliography{Biblio}

\end{document}